\documentclass{article}

\usepackage{microtype}
\usepackage{graphicx}
\usepackage{subcaption}
\usepackage{booktabs} 
\usepackage{pgfplots}
\pgfplotsset{compat=1.18}
\usepgfplotslibrary{fillbetween,groupplots}

\usepackage{hyperref}

\usepackage[preprint]{icml2026}

\usepackage{amsmath}
\usepackage{amssymb}
\usepackage{mathtools}
\usepackage{multirow}
\usepackage{amsthm}

\usepackage[capitalize,noabbrev]{cleveref}

\theoremstyle{plain}

\theoremstyle{definition}

\theoremstyle{remark}

\usepackage[textsize=tiny]{todonotes}

\icmltitlerunning{Uncertainty-aware Machine Learning Interatomic Potentials via Learned Functional Perturbations}

\begin{document}

\twocolumn[
  \icmltitle{Uncertainty-aware Machine Learning Interatomic Potentials \\ via Learned Functional Perturbations
}



  \icmlsetsymbol{equal}{*}
  
  \begin{icmlauthorlist}
    \icmlauthor{Olga Zaghen}{equal,yyy}
    \icmlauthor{Maksim Zhdanov}{equal,yyy}
    \icmlauthor{Dario Coscia}{xxx,yyy}
    \icmlauthor{David R. Wessels}{yyy}
    \icmlauthor{Erik J. Bekkers}{yyy}
  \end{icmlauthorlist}

  \icmlaffiliation{yyy}{AMLab, Universiy of Amsterdam}
  \icmlaffiliation{xxx}{mathLab, SISSA}

  \icmlcorrespondingauthor{Olga Zaghen}{o.zaghen@uva.nl}

  \icmlkeywords{Machine Learning, ICML}

  \vskip 0.3in
]



\printAffiliationsAndNotice{\icmlEqualContribution}

\begin{abstract}
Machine Learning Interatomic Potentials (MLIPs) achieve near ab initio accuracy at a
fraction of the cost of quantum-mechanical simulations, yet they remain prone to silent
failures on out-of-distribution configurations, making principled uncertainty
quantification (UQ) essential for error-aware simulations and active learning.
Existing non-ensemble UQ methods for MLIPs rely either on variational inference or on
parametric distributional assumptions, both of which add architectural complexity and
hyper-parameters that must be tuned per task.
Inspired by recent advances in probabilistic weather forecasting, we propose a simpler
alternative: turn a deterministic MLIP into a probabilistic one through learned functional
perturbations and finetune it end-to-end with the Continuous Ranked Probability Score (CRPS),
a proper scoring rule.
We validate the approach with an equivariant GNN (P-EGNN) trained from scratch and by finetuning the foundation model the Orb-v3 for silica.
On the N-body charged particle benchmark, P-EGNN improves CRPS over the state-of-the-art Bayesian MLIP method BLIP~\citep{coscia2025blips} by 19--32\% across all training sizes; on silica, P-Orb raises the Spearman correlation between predicted uncertainty and actual error from $0.75$ (BLIP-Orb) to $0.84$.

\end{abstract}

\section{Introduction}

Machine Learning Interatomic Potentials (MLIPs) have transformed simulation-based chemistry and materials science, enabling molecular dynamics at near ab-initio
accuracy with cost scaling linearly with system size~\citep{satorras2021n,batatia2022mace,neumann2024orb,wood2026family}.
Despite this progress, MLIPs are vulnerable to silent failures on out-of-distribution configurations, making principled uncertainty quantification (UQ) essential for error-aware simulations and active learning pipelines~\citep{kulichenko2023uncertainty,coscia2025blips}.

\textbf{The need for simpler UQ.}
Deep ensembles~\citep{lakshminarayanan2017simple} remain the de-facto standard for UQ in MLIPs~\citep{tan2023single}, yet their $K$-fold cost is prohibitive for foundation backbones with hundreds of millions of parameters where, in practice, only a single pretrained checkpoint is available~\citep{rhodes2025orb}. Single-model
alternatives such as MC Dropout~\citep{gal2016dropout}, variational Bayesian
potentials~\citep{coscia2025blips} or evidential regression~\citep{xu2025evidential} reduce
this cost, yet each adds its own machinery (inference networks, KL terms, higher-order
priors) that must be tuned per task.
In this work, we adopt a simpler approach inspired by probabilistic
weather forecasting~\citep{alet2025fgn, zhdanov2026sparse}: by perturbing network's weights, we convert a deterministic model into a probabilistic one and optimize it end-to-end with the Continuous Ranked
Probability Score (CRPS)~\citep{gneiting2007strictly}.

\textbf{Our contribution.}
We propose P-MLIP (Perturbed-MLIP), a lightweight plug-in module for training and fine-tuning MLIPs that employs learned functional perturbations~\citep{alet2025fgn} to introduce noise in the parameter space, yielding a probabilistic model with minimal overhead and no auxiliary networks or ad-hoc parameters.
42We evaluate P-MLIP in two settings: P-EGNN (EGNN from scratch on N-body) and P-Orb (Orb-v3 fine-tuned on silica). On N-body, P-EGNN improves CRPS over BLIP-EGNN by 19–32\% and is the only method whose spread-to-skill ratio is close to the calibrated value of $1$, at a fraction of the ensemble's training cost; on silica, P-Orb outperforms BLIP-Orb on all metrics at every training size, improving the Spearman correlation between predicted uncertainty and actual error from $0.75$ to $0.84$.

\section{Related Work}

\textbf{UQ methods for MLIPs.}
Most UQ methods for MLIPs fall into one of three families. \emph{Ensemble methods}
(deep ensembles~\citep{lakshminarayanan2017simple}, query-by-committee) remain the
empirical gold standard for both predictive accuracy and uncertainty in atomistic
models~\citep{tan2023single,kulichenko2023uncertainty}, but their $K$-fold inference
cost is prohibitive for large pretrained backbones~\citep{rhodes2025orb}.
\emph{Bayesian single-model methods} approximate the posterior over weights from a
single training run: MC Dropout~\citep{gal2016dropout} is cheap but typically
miscalibrated, while BLIP~\citep{coscia2025blips} (the closest prior work to ours) introduces
input-dependent variational dropout into MPNN weights, trained via an ELBO with an MSE
likelihood and a KL term against a weight prior. BLIP yields well-calibrated uncertainties
and improves accuracy over deterministic MLIPs, but requires a dedicated posterior
inference network and tuning of the KL coefficient. \emph{Evidential and conformal
approaches} place a higher-order distribution over predictions~\citep{soleimany2021evidential,xu2025evidential}
or wrap an existing model with distribution-free coverage guarantees via conformal
prediction~\citep{hu2024conformal}; both are single-pass but rely on either a parametric prior whose hyperparameters must be carefully tuned, or a held-out calibration set that yields prediction intervals rather than a full predictive distribution.

\textbf{CRPS-trained probabilistic forecasting.}
A line of recent work in machine-learning weather forecasting trains ensemble models
end-to-end with the CRPS as the probabilistic objective. AIFS-CRPS~\citep{lang2024aifs}
replaces a deterministic loss with CRPS on a graph-transformer backbone, while
FGN~\citep{alet2025fgn} and \textsc{Mosaic}~\citep{zhdanov2026sparse} obtain
the predictive ensemble by perturbing the model's weights.
A common design across these methods is that the noise is shared across all spatial
locations within a forward pass, producing globally consistent perturbations that
reflect coherent physical uncertainty.

\section{Method}

\subsection{Noise Injection Architecture}

Let $f_\theta$ be an MLIP implemented as a $L$-layer MPNN with learnable parameters $\theta$, where each
layer applies the message and node update functions parameterized by MLPs. We introduce
a global noise vector $\mathbf{z}\!\in\!\mathbb{R}^{d_z}$:
\begin{equation}
    \mathbf{z} \sim \mathcal{N}\!\left(\mathbf{0},\, W_z W_z^\top\right),
    \label{eq:noise_gen}
\end{equation}
where $W_z\!\in\!\mathbb{R}^{d_z\times d_z}$ is a learnable linear transformation.
For each forward pass, a new independent $\mathbf{z}$ is generated,
ensuring different samples see different noise
realizations at inference time.

\textbf{Injection into MLP blocks.}
Each MPNN layer $l\!\in\!\{1,\ldots,L\}$ contains two MLPs that we perturb: the message function $M_l$ (edge update) and the node update function $U_l$. Each of these is a small MLP composed of $S$ stacked linear sub-layers with activations in between; we call one such MLP a \emph{block} and index the $B\!=\!2L$ perturbed blocks by $b$. Within each block $b$, let $s\!=\!1,\ldots,S$ index its linear sub-layers and let $\mathbf{x}_b$ denote the block input. We inject $\mathbf{z}$ \emph{only} at the first sub-layer of each block,
\begin{equation}
  \mathbf{h}^{(1)}_b \;=\; W^{(1)}_b\,\mathbf{x}_b \;+\; W^{\mathrm{noise}}_b\,\mathbf{z},
  \label{eq:inject}
\end{equation}
while sub-layers $s\!\geq\!2$ follow the standard MLP composition without any further noise term. Each $W^{\mathrm{noise}}_b\!\in\!\mathbb{R}^{d_h\times d_z}$ is a distinct learnable matrix initialised to \emph{zero}. Injecting at $s\!=\!1$ keeps each block's noise contribution structurally a linear functional perturbation of the block's raw input, as classical additive input noise \citep{bishop1995training,graves2013speech}. Zero-initialisation makes the noise contribution $W^{\mathrm{noise}}_b\mathbf{z}$ vanish at $t\!=\!0$ regardless of $\mathbf{z}$, so the model boots up as the deterministic backbone, which is essential when fine-tuning a pretrained foundation model (P-Orb) and removes the need for noise schedules or KL warm-ups when training from scratch (P-EGNN). The noise channel is then activated progressively as the CRPS objective rewards predictive spread. Position-update MLPs are deliberately excluded from the perturbed blocks to preserve equivariance.

\textbf{Graph-level noise.}
Critically, the same noise vector $\mathbf{z}$ is broadcast to \emph{all atoms} within a molecular graph. This produces a globally consistent perturbation of the potential energy surface, analogous to the learned functional perturbations used by~\citet{alet2025fgn} and \citet{zhdanov2026sparse}, rather than independent per-atom stochasticity.


\textbf{Model prediction and uncertainty.}
Both at training and at inference, $K$ independent forward passes are executed with independently drawn
$\mathbf{z}^{(1)},\ldots,\mathbf{z}^{(K)}$, yielding $K$ output samples
$\{\hat{\mathbf{y}}^{(k)}\}_{k=1}^K$. The number of samples $K$ may differ between training, validation, and inference.

\textbf{P-EGNN.}
We apply the above to the E($n$)-Equivariant Graph Neural Network
(EGNN;~\citealp{satorras2021n}), perturbing all MLP blocks on the edge (aggregation) and node
update functions (Eq.~\ref{eq:inject}). Position update networks
are left noise-free to avoid equivariance breaking.

\textbf{P-Orb.}
For the silica experiments, we wrap the pretrained Orb-v3 model~\citep{rhodes2025orb}
with noise hooks inserted at the first linear layer of each GNN stack's node and edge
MLPs via PyTorch forward hooks, without modifying the base architecture. Zero
initialization keeps the model identical to the pretrained Orb-v3 at the start of
fine-tuning.

\subsection{Training Objective: Fair CRPS}

We train P-MLIP by minimizing the \emph{unbiased} (fair) estimator of the Continuous Ranked Probability Score~\citep{zamo2018estimation}.

\textbf{Univariate estimator.}
For a scalar target $y\!\in\!\mathbb{R}$ with $K$ ensemble samples $\hat{y}^{1:K}$, the fair CRPS estimator is
\begin{equation}
\begin{split}
  \mathrm{CRPS}\bigl(\hat{y}^{1:K},\,y\bigr)
  &= \frac{1}{K}\!\sum_{k=1}^{K}\!\bigl|\hat{y}^{(k)}\!-\!y\bigr| \\
  &\quad - \frac{1}{2K(K{-}1)}\!\sum_{i\neq j}\!\bigl|\hat{y}^{(i)}\!-\!\hat{y}^{(j)}\bigr|,
\end{split}
\label{eq:crps_uni}
\end{equation}
where the first (\emph{reliability}) term penalizes deviation from ground truth and the second (\emph{sharpness}) term rewards ensemble spread. Excluding the diagonal $i\!=\!j$ (whose contribution vanishes deterministically) gives the unbiased estimator that prevents the degenerate point-ensemble optimum~\citep{zamo2018estimation}. With $K\!=\!1$, Eq.~\ref{eq:crps_uni} reduces to the absolute error.

\textbf{Extension to per-atom and per-structure targets.}
MLIP outputs are not scalars: a structure with $N$ atoms carries a force matrix $\mathbf{F}\!\in\!\mathbb{R}^{N\times 3}$ and a single energy $E\!\in\!\mathbb{R}$. Reading the target as a vector $\mathbf{y}\!\in\!\mathbb{R}^D$ with $D\!=\!1$ for an energy and $D\!=\!3N$ for a force matrix, we apply the scalar fair-CRPS estimator (Eq.~\ref{eq:crps_uni}) \emph{to every entry} and average:
\begin{equation}
\begin{split}
  \mathcal{L}_{\mathrm{CRPS}}\bigl(\hat{\mathbf{y}}^{1:K},\,\mathbf{y}\bigr)
  &= \frac{1}{D}\!\sum_{d=1}^{D}\mathrm{CRPS}\bigl(\hat{y}_d^{1:K},\, y_d\bigr) \\
  &= \frac{1}{D K}\!\sum_{k,d}\!\bigl|\hat{y}^{(k)}_d\!-\!y_d\bigr| \\
  &\quad - \frac{1}{2 D K(K{-}1)}\!\sum_{\substack{i\neq j\\ d}}\!\bigl|\hat{y}^{(i)}_d\!-\!\hat{y}^{(j)}_d\bigr|,
\end{split}
\label{eq:crps}
\end{equation}
where $\mathbf{y}$ stands for either an energy scalar $E$ or a force matrix $\mathbf{F}$. Swapping the order of the sums over samples and entries, Eq.~\ref{eq:crps} is exactly $1/D$ times the multivariate $\ell^{1}$ energy score~\citep{gneiting2007strictly}, a proper scoring rule for multivariate targets. The constant rescaling is optimisationally inert, but the per-entry decomposition is what we use in practice: it makes the unbiasedness of the scalar fair estimator manifest at every $d$ without introducing cross-entry covariance terms, and the explicit $1/D$ prefactor (rather than the bare $\ell^{1}$ \emph{sum}, which would scale with $3N$ for forces) brings energy and forces onto comparable per-entry magnitudes without separate loss weights. With $K\!=\!1$, Eq.~\ref{eq:crps} reduces to the mean absolute error.

For P-Orb we follow the joint energy/forces training convention of BLIP~\citep{coscia2025blips} and minimize the sum
$\mathcal{L}_{\mathrm{CRPS}}(E/N) + \mathcal{L}_{\mathrm{CRPS}}(\mathbf{F})$ of
a per-atom energy CRPS (structure-level energy is divided by $N$ atoms, so energy and forces train on comparable scales) and a per-structure forces CRPS, both computed via
Eq.~\ref{eq:crps}.

\subsection{Connection with Bayesian Methods}
In a Bayesian neural network (BNN), predictive uncertainty arises from the posterior $p(\theta \mid \mathcal{D})$ over the network parameters, which concentrates with more data and captures reducible epistemic uncertainty.
P-MLIP performs no posterior inference: the parameters $\theta$ are point estimates, and ensemble spread originates from sampling $\mathbf{z}$ from the learned prior $p(\mathbf{z})$ rather than from parameter uncertainty.
This distinction means the two approaches are complementary: one could, in principle, combine BLIP's variational inference with a P-MLIP backbone to capture both epistemic weight uncertainty and the learned noise structure simultaneously.

\subsection{Implementation Remarks}

\textbf{Equivariance.}
For P-EGNN, $\mathbf{z}$ is injected only into the invariant scalar MLP blocks (edge and node update functions, which operate on pairwise distances and node features); position update networks are left noise-free. As a result, for any fixed noise realization $\mathbf{z}$, the model $f_\theta(\cdot,\mathbf{z})$ is exactly E($n$)-equivariant: $f_\theta(g\!\cdot\!\mathbf{x},\mathbf{z}) = g\!\cdot\! f_\theta(\mathbf{x},\mathbf{z})$
for all $g\!\in\!\mathrm{E}(n)$. This is strictly stronger than the distributional guarantee (equivariance of the mean) that BLIP achieves through a dedicated invariant inference network~\citep{coscia2025blips}; P-EGNN obtains the same symmetry for free, since the noise prior $p(\mathbf{z})$ is defined independently of the molecular geometry.
For P-Orb, the same per-sample equivariance holds by construction, as the noise hooks target only the invariant scalar channels (node and edge MLPs) of the Orb-v3 architecture.



\begin{figure*}[t]
  \centering
  \begin{minipage}[t]{0.49\textwidth}
    \centering
    \resizebox{\linewidth}{!}{\definecolor{blipblue}{RGB}{31,119,180}
\definecolor{pegnnred}{RGB}{214,96,77}

\begin{tikzpicture}
\begin{groupplot}[
  group style={
    group size=3 by 1,
    horizontal sep=0.7cm,
  },
  width=0.36\linewidth,
  height=0.36\linewidth,
  xmode=log,
  xtick={1000,3000,10000},
  xticklabels={$1\mathrm{k}$,$3\mathrm{k}$,$10\mathrm{k}$},
  xlabel style={font=\scriptsize, yshift=2pt},
  tick label style={font=\scriptsize},
  legend style={
    font=\scriptsize,
    draw=gray!60,
    fill=white,
    fill opacity=0.85,
    text opacity=1,
    inner sep=2pt,
    column sep=0.5em,
    legend cell align=left,
  },
  legend columns=-1,
  ymajorgrids=true,
  grid style={dotted, gray!40},
  every axis plot/.append style={line width=1.5pt},
  clip=false,
]

\nextgroupplot[
  legend to name=nbodylegend,
  ymin=3.5, ymax=9.5,
]
\addplot[name path=blip_mse_up, draw=none, forget plot]
  coordinates {(1000,8.14) (3000,6.88) (10000,4.98)};
\addplot[name path=blip_mse_lo, draw=none, forget plot]
  coordinates {(1000,6.64) (3000,5.38) (10000,4.16)};
\addplot[blipblue!35, fill opacity=0.8, forget plot]
  fill between[of=blip_mse_up and blip_mse_lo];
\addplot[name path=pegnn_mse_up, draw=none, forget plot]
  coordinates {(1000,8.49) (3000,6.47) (10000,4.36)};
\addplot[name path=pegnn_mse_lo, draw=none, forget plot]
  coordinates {(1000,7.29) (3000,5.71) (10000,4.02)};
\addplot[pegnnred!35, fill opacity=0.8, forget plot]
  fill between[of=pegnn_mse_up and pegnn_mse_lo];
\addplot[color=blipblue, mark=*, mark size=2.5pt,
  mark options={draw=black, fill=blipblue, line width=1.0pt}]
  coordinates {(1000,7.39) (3000,6.13) (10000,4.57)};
\addlegendentry{BLIP}
\addplot[color=pegnnred, mark=*, mark size=2.5pt,
  mark options={draw=black, fill=pegnnred, line width=1.0pt}]
  coordinates {(1000,7.89) (3000,6.09) (10000,4.19)};
\addlegendentry{P-EGNN (Ours)}

\nextgroupplot[
  xlabel={Training set size $n$},
  ymin=1.0, ymax=3.8,
]
\addplot[name path=blip_crps_up, draw=none, forget plot]
  coordinates {(1000,3.23) (3000,2.47) (10000,2.19)};
\addplot[name path=blip_crps_lo, draw=none, forget plot]
  coordinates {(1000,2.49) (3000,2.23) (10000,1.81)};
\addplot[blipblue!35, fill opacity=0.8, forget plot]
  fill between[of=blip_crps_up and blip_crps_lo];
\addplot[name path=pegnn_crps_up, draw=none, forget plot]
  coordinates {(1000,2.47) (3000,1.84) (10000,1.43)};
\addplot[name path=pegnn_crps_lo, draw=none, forget plot]
  coordinates {(1000,2.15) (3000,1.68) (10000,1.29)};
\addplot[pegnnred!35, fill opacity=0.8, forget plot]
  fill between[of=pegnn_crps_up and pegnn_crps_lo];
\addplot[color=blipblue, mark=*, mark size=2.5pt,
  mark options={draw=black, fill=blipblue, line width=1.0pt}, forget plot]
  coordinates {(1000,2.86) (3000,2.35) (10000,2.00)};
\addplot[color=pegnnred, mark=*, mark size=2.5pt,
  mark options={draw=black, fill=pegnnred, line width=1.0pt}, forget plot]
  coordinates {(1000,2.31) (3000,1.76) (10000,1.36)};

\nextgroupplot[
  ymin=0.25, ymax=1.1,
]
\addplot[black, dashed, line width=0.8pt, forget plot]
  coordinates {(1000,1.0) (10000,1.0)};
\addplot[name path=blip_ssr_up, draw=none, forget plot]
  coordinates {(1000,0.437) (3000,0.453) (10000,0.441)};
\addplot[name path=blip_ssr_lo, draw=none, forget plot]
  coordinates {(1000,0.363) (3000,0.389) (10000,0.397)};
\addplot[blipblue!35, fill opacity=0.8, forget plot]
  fill between[of=blip_ssr_up and blip_ssr_lo];
\addplot[name path=pegnn_ssr_up, draw=none, forget plot]
  coordinates {(1000,0.829) (3000,0.948) (10000,0.928)};
\addplot[name path=pegnn_ssr_lo, draw=none, forget plot]
  coordinates {(1000,0.665) (3000,0.712) (10000,0.824)};
\addplot[pegnnred!35, fill opacity=0.8, forget plot]
  fill between[of=pegnn_ssr_up and pegnn_ssr_lo];
\addplot[color=blipblue, mark=*, mark size=2.5pt,
  mark options={draw=black, fill=blipblue, line width=1.0pt}, forget plot]
  coordinates {(1000,0.400) (3000,0.421) (10000,0.419)};
\addplot[color=pegnnred, mark=*, mark size=2.5pt,
  mark options={draw=black, fill=pegnnred, line width=1.0pt}, forget plot]
  coordinates {(1000,0.747) (3000,0.830) (10000,0.876)};

\end{groupplot}

\node[anchor=south, font=\scriptsize] at ([yshift=4pt]group c1r1.north)
  {MSE{\scalebox{0.55}{$\!\times\!10^{-3}$}} $\downarrow$};
\node[anchor=south, font=\scriptsize] at ([yshift=4pt]group c2r1.north)
  {CRPS{\scalebox{0.55}{$\!\times\!10^{-2}$}} $\downarrow$};
\node[anchor=south, font=\scriptsize] at ([yshift=4pt]group c3r1.north)
  {SSR $\uparrow$ (ideal $= 1$)};

\node[anchor=south] at ([yshift=18pt]group c2r1.north) {%
  \pgfplotslegendfromname{nbodylegend}};

\end{tikzpicture}}
    \caption{N-body test performance vs.\ training size (mean~$\pm$~std over 4 seeds, shaded). Left: MSE; Center: CRPS; Right: spread-to-skill ratio SSR. P-EGNN consistently achieves the best CRPS and the SSR closest to $1$ at every training size, with the calibration gap widening as $n$ grows.}
    \label{fig:nbody}
  \end{minipage}\hfill
  \begin{minipage}[t]{0.49\textwidth}
    \centering
    \resizebox{\linewidth}{!}{
\definecolor{blipblue}{RGB}{31,119,180}
\definecolor{porbred}{RGB}{214,96,77}

\begin{tikzpicture}
\begin{groupplot}[
  group style={
    group size=3 by 1,
    horizontal sep=0.7cm,
  },
  width=0.36\linewidth,
  height=0.36\linewidth,
  xmode=log,
  xtick={32,128,1024},
  xticklabels={$32$,$128$,$1024$},
  xlabel style={font=\scriptsize, yshift=2pt},
  tick label style={font=\scriptsize},
  legend style={
    font=\scriptsize,
    draw=gray!60,
    fill=white,
    fill opacity=0.85,
    text opacity=1,
    inner sep=2pt,
    column sep=0.5em,
    legend cell align=left,
  },
  legend columns=-1,
  ymajorgrids=true,
  grid style={dotted, gray!40},
  every axis plot/.append style={line width=1.5pt},
  clip=false,
]

\nextgroupplot[
  legend to name=silicalegend,
  ymin=0.5, ymax=7.5,
]
\addplot[name path=blip_mae_up, draw=none, forget plot]
  coordinates {(32,6.09) (128,3.40) (1024,1.61)};
\addplot[name path=blip_mae_lo, draw=none, forget plot]
  coordinates {(32,6.01) (128,3.14) (1024,1.61)};
\addplot[blipblue!35, fill opacity=0.8, forget plot]
  fill between[of=blip_mae_up and blip_mae_lo];
\addplot[name path=porb_mae_up, draw=none, forget plot]
  coordinates {(32,2.52) (128,2.00) (1024,1.43)};
\addplot[name path=porb_mae_lo, draw=none, forget plot]
  coordinates {(32,2.26) (128,1.72) (1024,1.33)};
\addplot[porbred!35, fill opacity=0.8, forget plot]
  fill between[of=porb_mae_up and porb_mae_lo];
\addplot[color=blipblue, mark=*, mark size=2.5pt,
  mark options={draw=black, fill=blipblue, line width=1.0pt}]
  coordinates {(32,6.05) (128,3.27) (1024,1.61)};
\addlegendentry{BLIP}
\addplot[color=porbred, mark=*, mark size=2.5pt,
  mark options={draw=black, fill=porbred, line width=1.0pt}]
  coordinates {(32,2.39) (128,1.86) (1024,1.38)};
\addlegendentry{P-ORB (Ours)}

\nextgroupplot[
  xlabel={Training set size $n$},
  ymin=0.5, ymax=6.5,
]
\addplot[name path=blip_crps_up, draw=none, forget plot]
  coordinates {(32,5.82) (128,3.00) (1024,1.17)};
\addplot[name path=blip_crps_lo, draw=none, forget plot]
  coordinates {(32,5.74) (128,2.60) (1024,1.15)};
\addplot[blipblue!35, fill opacity=0.8, forget plot]
  fill between[of=blip_crps_up and blip_crps_lo];
\addplot[name path=porb_crps_up, draw=none, forget plot]
  coordinates {(32,2.09) (128,1.57) (1024,0.968)};
\addplot[name path=porb_crps_lo, draw=none, forget plot]
  coordinates {(32,1.79) (128,1.39) (1024,0.948)};
\addplot[porbred!35, fill opacity=0.8, forget plot]
  fill between[of=porb_crps_up and porb_crps_lo];
\addplot[color=blipblue, mark=*, mark size=2.5pt,
  mark options={draw=black, fill=blipblue, line width=1.0pt}, forget plot]
  coordinates {(32,5.78) (128,2.80) (1024,1.16)};
\addplot[color=porbred, mark=*, mark size=2.5pt,
  mark options={draw=black, fill=porbred, line width=1.0pt}, forget plot]
  coordinates {(32,1.94) (128,1.48) (1024,0.958)};

\nextgroupplot[
  ymin=0.2, ymax=1.05,
]
\addplot[name path=blip_spear_up, draw=none, forget plot]
  coordinates {(32,0.412) (128,0.534) (1024,0.755)};
\addplot[name path=blip_spear_lo, draw=none, forget plot]
  coordinates {(32,0.406) (128,0.398) (1024,0.747)};
\addplot[blipblue!35, fill opacity=0.8, forget plot]
  fill between[of=blip_spear_up and blip_spear_lo];
\addplot[name path=porb_spear_up, draw=none, forget plot]
  coordinates {(32,0.882) (128,0.889) (1024,0.852)};
\addplot[name path=porb_spear_lo, draw=none, forget plot]
  coordinates {(32,0.828) (128,0.801) (1024,0.820)};
\addplot[porbred!35, fill opacity=0.8, forget plot]
  fill between[of=porb_spear_up and porb_spear_lo];
\addplot[black, dashed, line width=0.8pt, forget plot]
  coordinates {(32,1.0) (1024,1.0)};
\addplot[color=blipblue, mark=*, mark size=2.5pt,
  mark options={draw=black, fill=blipblue, line width=1.0pt}, forget plot]
  coordinates {(32,0.409) (128,0.466) (1024,0.751)};
\addplot[color=porbred, mark=*, mark size=2.5pt,
  mark options={draw=black, fill=porbred, line width=1.0pt}, forget plot]
  coordinates {(32,0.855) (128,0.845) (1024,0.836)};

\end{groupplot}

\node[anchor=south, font=\scriptsize] at ([yshift=4pt]group c1r1.north)
  {F-MAE{\scalebox{0.55}{[meV/\AA]}} $\downarrow$};
\node[anchor=south, font=\scriptsize] at ([yshift=4pt]group c2r1.north)
  {F-CRPS{\scalebox{0.55}{[meV/\AA]}} $\downarrow$};
\node[anchor=south, font=\scriptsize] at ([yshift=4pt]group c3r1.north)
  {F-Spear $\uparrow$ (ideal $= 1$)};

\node[anchor=south] at ([yshift=18pt]group c2r1.north) {%
  \pgfplotslegendfromname{silicalegend}};

\end{tikzpicture}}
    \caption{Silica forces performance vs.\ training size (mean~$\pm$~std over 4 seeds, shaded). Left: F-MAE; Center: F-CRPS; Right: Spearman~$\rho$ (higher is better). P-Orb outperforms BLIP on all metrics at every training size.}
    \label{fig:silica}
  \end{minipage}
\end{figure*}

\begin{table*}[t]
  \centering
  \begin{minipage}[t]{0.49\textwidth}
    \centering
    \caption{N-body test performance at $n\!=\!3000$ (mean $\pm$ std over 4 seeds).
      MSE $\times10^{-3}$; CRPS $\times10^{-2}$ (lower is better for both);
      SSR: spread-to-skill ratio with finite-$K$ correction, $K\!=\!100$ for BLIP/P-EGNN and $K\!=\!3$ for the deep ensemble.
      Dashes: metric not applicable. \textbf{Bold}: best per column.
      $^\dagger$std underestimates true variance (LOO splits share members).}
    \label{tab:nbody}
    \resizebox{\linewidth}{!}{%
      \begin{tabular}{lccc}
        \toprule
        Model & MSE ↓ & CRPS ↓ & SSR ↑ \\
        \midrule
        EGNN & $6.83\pm0.31$ & — & — \\
        BLIP & $6.13\pm0.75$ & $2.35\pm0.12$ & $0.42\pm0.03$ \\
        EGNN ens.$^\dagger$ & $\mathbf{5.84\pm0.13}$ & $1.88\pm0.10$ & $0.58\pm0.02$ \\
        P-EGNN \textbf{(Ours)} & $6.09\pm0.38$ & $\mathbf{1.76\pm0.08}$ & $\mathbf{0.83\pm0.12}$ \\
        \bottomrule
      \end{tabular}%
    }
  \end{minipage}\hfill
  \begin{minipage}[t]{0.49\textwidth}
    \centering
    \caption{Silica forces results at $n\!=\!1024$ training structures
      (mean~$\pm$~std, 4 seeds). F-MAE in meV/\AA;
      F-CRPS: forces CRPS (lower is better);
      F-Spear: Spearman $\rho$ with squared error (higher is better).
      \textbf{Bold}: best per metric.}
    \label{tab:silica}
    \resizebox{\linewidth}{!}{%
      \begin{tabular}{lccc}
        \toprule
        Model & F-MAE ↓ & F-CRPS ↓ & F-Spear ↑ \\
        \midrule
        BLIP-Orb & $1.61\pm0.00$ & $1.16\pm0.01$ & $0.75\pm0.04$ \\
        P-Orb \textbf{(Ours)} & $\mathbf{1.38\pm0.05}$ & $\mathbf{0.96\pm0.10}$ & $\mathbf{0.84\pm0.16}$ \\
        \bottomrule
      \end{tabular}%
    }
  \end{minipage}
\end{table*}

\textbf{Computational overhead.}
P-MLIP and BLIP share the same Monte Carlo structure at inference: producing a full predictive distribution requires $K$ forward passes through the noisy backbone, matching the cost of a $K$-member deep ensemble but from a \emph{single} trained model. Both methods can also produce a one-pass MAP-style point prediction by zeroing the stochastic component ($\mathbf{z}\!=\!\mathbf{0}$ in P-MLIP, $\boldsymbol{\omega}\!=\!\boldsymbol{\theta}$ in BLIP).
The asymmetry between the two methods is at \emph{training} time: BLIP's ELBO uses a single MC sample per step ($S\!=\!1$), whereas the fair-CRPS loss (Eq.~\ref{eq:crps}) requires $K_{\mathrm{train}}\!\geq\!2$ and we default to $K_{\mathrm{train}}\!=\!10$, so each P-MLIP gradient step performs $K_{\mathrm{train}}\!\times$ more forward+backward passes than a BLIP step. Against this, P-MLIP has no inference network to train and no KL coefficient to tune. In the EGNN case, the additional noise-projection parameters introduced by P-EGNN (one $W_z\!\in\!\mathbb{R}^{d_z\times d_z}$ shared across the model plus one $W^{\mathrm{noise}}_b\!\in\!\mathbb{R}^{d_h\times d_z}$ per perturbed block) amount to ${\sim}13\%$ of the EGNN backbone (\cref{app:nbody_params}) and are independent of the inference ensemble size $K$.

\section{Experiments}

\subsection{N-body: Training from Scratch}

\textbf{Setup.}
We evaluate on the Coloumbs’ N-body particle benchmark \citep{fuchs2020se}: the task is to predict particle positions after 1000 time steps given initial positions and velocities. We vary the training set size $n\!\in\!\{1000, 3000, 10000\}$ and report test MSE (on the sample mean), CRPS (Eq.~\ref{eq:crps}), and the spread-to-skill ratio (SSR); SSR$\,=\,1$ corresponds to a calibrated ensemble, SSR$\,<\,1$ to under-dispersion (over-confidence) and SSR$\,>\,1$ to under-confidence. All metrics are averaged over 4 random seeds.

\textbf{Baselines.}
We compare P-EGNN against three baselines using the same EGNN backbone (4 layers,
64 hidden units):
(i)~EGNN~\citep{satorras2021n}: deterministic, trained with MSE;
(ii)~BLIP(-EGNN)~\citep{coscia2025blips}, with the BLIP method applied to all MLP
layers, trained with MSE + KL;
(iii)~EGNN-ensemble: a deep ensemble of 3 independently trained EGNN models. P-EGNN uses $K\!=\!10$ samples during training, $K\!=\!50$ at validation, and $K\!=\!100$ at test time; $d_z\!=\!32$. Importantly, P-EGNN adds only ${\sim}13\%$ parameters over a single EGNN backbone, whereas the $K\!=\!3$ deep ensemble triples the parameter count; see \cref{app:nbody_params} for the full breakdown.

\textbf{Results.}
\cref{fig:nbody} reports scaling across all three metrics, and \cref{tab:nbody} gives the head-to-head snapshot at $n\!=\!3000$ including the deep ensemble. P-EGNN obtains the best CRPS at every training size, with the gap over BLIP widening from 19\% at $n\!=\!1000$ to 32\% at $n\!=\!10000$, and edging out the deep ensemble at $n\!=\!3000$ at a fraction of its training cost. SSR tells the same story on calibration: P-EGNN reaches $0.83\!\pm\!0.12$ at $n\!=\!3000$ and trends towards $1$ as $n$ grows ($0.75\!\to\!0.88$), while BLIP is severely under-dispersed across all training sizes ($\mathrm{SSR}\!\approx\!0.42$, roughly half of P-EGNN's), and the $K\!=\!3$ deep ensemble is also under-dispersed ($0.58\!\pm\!0.02$) as expected at small $K$. On MSE, CRPS training is competitive despite never optimizing squared error: P-EGNN matches or surpasses BLIP from $n\!=\!3000$ onward and sits within one standard deviation of the ensemble's best.

\subsection{Silica: Pretrained Foundation Model}

\textbf{Setup.}
We apply P-Orb to the silica glass (SiO$_2$, 699 atoms per structure) benchmark
from \citet{coscia2025blips}, fine-tuning a pretrained Orb-v3 model~\citep{rhodes2025orb}.
The baseline BLIP-Orb applies BLIP to the same foundation model.
Training sizes span $n\!\in\!\{32, 128, 1024\}$; results are averaged over 4 seeds.
We report forces MAE (meV/\AA), forces CRPS (F-CRPS), and Spearman rank correlation
$\rho$ between per-structure squared error and predicted variance.
Energy results show high variance across seeds and are deferred to future work.

\textbf{Results.}
\cref{fig:silica} reports scaling across all three metrics, and \cref{tab:silica} gives the head-to-head snapshot at $n\!=\!1024$. P-Orb obtains the best F-CRPS at every training size, with the gap over BLIP-Orb largest in the data-scarce regime (3$\times$ at $n\!=\!32$) and persisting even at $n\!=\!1024$ ($0.96$ vs.\ $1.16$ meV/\AA). F-MAE follows the same ordering: CRPS training improves the mean prediction without ever optimizing squared error, with the advantage again most pronounced at small $n$ and narrowing but not closing as data grows ($1.38$ vs.\ $1.61$ at $n\!=\!1024$).
On uncertainty quality, P-Orb's Spearman $\rho$ is high and stable across training sizes ($\geq\!0.84$), whereas BLIP-Orb only catches up at $n\!=\!1024$ ($0.75$); at smaller $n$ its ensemble spread is poorly correlated with actual errors, consistent with posterior collapse on this fine-tuning regime.

\section{Discussion and Conclusion}

We presented P-MLIP, a simple, architecture-agnostic approach to uncertainty quantification in MLIPs via learned noise injection and CRPS training, requiring no variational inference and adding no structural complexity beyond a zero-initialized noise projection. Empirically, on the N-body benchmark P-EGNN improves CRPS over BLIP-EGNN by 19--32\% across training sizes with a spread-to-skill ratio close to the calibrated value of $1$ where both BLIP and a small deep ensemble are markedly under-dispersed, and also obtains the best MSE at $n\!\geq\!3000$ despite never minimizing squared error directly, consistent with CRPS training implicitly regularizing the ensemble mean as a by-product of calibration. On silica, P-Orb's Spearman $\rho$ substantially exceeds BLIP-Orb's at every training size, confirming that its ensemble spread is predictive of actual errors, a property especially important for active learning where acquisition decisions hinge on reliable uncertainty. Taken together, these results indicate that the simpler design of P-MLIP (fixed noise through a learnable projection optimized directly with CRPS, rather than input-dependent perturbations trained via MSE + KL) consistently yields better calibration-sensitive metrics across all experiments. As a positive byproduct, the same design also gives per-sample E($n$)-equivariance for free, strictly stronger than the distributional symmetry obtained through BLIP's invariant inference network, since the noise prior is geometry-independent. A few caveats remain: higher seed-to-seed variance in P-Orb energy predictions on silica (sensitivity to loss balancing between energy and forces CRPS terms), a noise structure bounded by a single linear projection, and the $K$ forward passes at inference shared with all ensemble methods. Natural extensions are integration into active-learning loops and scaling the noise injection to larger equivariant architectures.

\section*{Impact Statement}

This paper presents work whose goal is to advance the field of Machine Learning
for computational chemistry and materials science. There are many potential societal
consequences of our work, none which we feel must be specifically highlighted here.

\section{Acknowledgments}
OZ acknowledges her funding from the CaLiForNIA project (Marie Skłodowska-Curie Actions Doctoral
Network 2022). MZ acknowledges support from Microsoft Research
AI4Science. DW is funded by a public grant of the Dutch Cancer Society (KWF) under subsidy (15059/2022-PPS2). This work used the Dutch national
e-infrastructure with the support of the SURF Cooperative. Computations were partially
performed using the University of Amsterdam - Science
Faculty (UvA/FNWI) HPC Facility.

\bibliography{example_paper}

@article{coscia2025blips,
  title={BLIPs: Bayesian Learned Interatomic Potentials},
  author={Coscia, Dario and de Haan, Pim and Welling, Max},
  journal={arXiv preprint arXiv:2508.14022},
  year={2025}
}

@article{zhdanov2026sparse,
  title={(Sparse) Attention to the Details: Preserving Spectral Fidelity in ML-based Weather Forecasting Models},
  author={Zhdanov, Maksim and Lucic, Ana and Welling, Max and van de Meent, Jan-Willem},
  journal={arXiv preprint arXiv:2604.16429},
  year={2026}
}

@inproceedings{satorras2021n,
  title={E (n) equivariant graph neural networks},
  author={Satorras, V{\i}ctor Garcia and Hoogeboom, Emiel and Welling, Max},
  booktitle={International conference on machine learning},
  pages={9323--9332},
  year={2021},
  organization={PMLR}
}

@article{lakshminarayanan2017simple,
  title={Simple and scalable predictive uncertainty estimation using deep ensembles},
  author={Lakshminarayanan, Balaji and Pritzel, Alexander and Blundell, Charles},
  journal={Advances in neural information processing systems},
  volume={30},
  year={2017}
}

@inproceedings{gal2016dropout,
  title={Dropout as a bayesian approximation: Representing model uncertainty in deep learning},
  author={Gal, Yarin and Ghahramani, Zoubin},
  booktitle={international conference on machine learning},
  pages={1050--1059},
  year={2016},
  organization={PMLR}
}

@article{batatia2022mace,
  title={MACE: Higher order equivariant message passing neural networks for fast and accurate force fields},
  author={Batatia, Ilyes and Kovacs, David P and Simm, Gregor and Ortner, Christoph and Cs{\'a}nyi, G{\'a}bor},
  journal={Advances in neural information processing systems},
  volume={35},
  pages={11423--11436},
  year={2022}
}

@article{rhodes2025orb,
  title={Orb-v3: atomistic simulation at scale},
  author={Rhodes, Benjamin and Vandenhaute, Sander and {\v{S}}imkus, Vaidotas and Gin, James and Godwin, Jonathan and Duignan, Tim and Neumann, Mark},
  journal={arXiv preprint arXiv:2504.06231},
  year={2025}
}

@article{zamo2018estimation,
  title={Estimation of the continuous ranked probability score with limited information and applications to ensemble weather forecasts},
  author={Zamo, Micha{\"e}l and Naveau, Philippe},
  journal={Mathematical Geosciences},
  volume={50},
  number={2},
  pages={209--234},
  year={2018},
  publisher={Springer}
}

@article{kulichenko2023uncertainty,
  title={Uncertainty-driven dynamics for active learning of interatomic potentials},
  author={Kulichenko, Maksim and Barros, Kipton and Lubbers, Nicholas and Li, Ying Wai and Messerly, Richard and Tretiak, Sergei and Smith, Justin S and Nebgen, Benjamin},
  journal={Nature computational science},
  volume={3},
  number={3},
  pages={230--239},
  year={2023},
  publisher={Nature Publishing Group US New York}
}

@article{gneiting2007strictly,
  title={Strictly proper scoring rules, prediction, and estimation},
  author={Gneiting, Tilmann and Raftery, Adrian E},
  journal={Journal of the American statistical Association},
  volume={102},
  number={477},
  pages={359--378},
  year={2007},
  publisher={Taylor \& Francis}
}

@article{alet2025fgn,
  title={Skillful joint probabilistic weather forecasting from marginals},
  author={Alet, Ferran and Price, Ilan and El-Kadi, Andrew and Masters, Dominic and Markou, Stratis and Andersson, Tom R and Stott, Jacklynn and Lam, Remi and Willson, Matthew and Sanchez-Gonzalez, Alvaro and Battaglia, Peter},
  journal={arXiv preprint arXiv:2506.10772},
  year={2025}
}

@article{lang2024aifs,
  title={AIFS-CRPS: Ensemble forecasting using a model trained with a loss function based on the {C}ontinuous {R}anked {P}robability {S}core},
  author={Lang, Simon and Alexe, Mihai and Clare, Mariana C A and Roberts, Christopher and Adewoyin, Rilwan and Bouallegue, Zied Ben and Chantry, Matthew and Dramsch, Jesper and Dueben, Peter D and Hahner, Sara and others},
  journal={arXiv preprint arXiv:2412.15832},
  year={2024}
}

@article{xu2025evidential,
  title={Evidential deep learning for interatomic potentials},
  author={Xu, Han and Cui, Taoyong and Tang, Chenyu and Ma, Jinzhe and Zhou, Dongzhan and Li, Yuqiang and Gao, Xiang and Gong, Xingao and Ouyang, Wanli and Zhang, Shufei and others},
  journal={Nature Communications},
  year={2025},
  publisher={Nature Publishing Group UK London}
}

@article{hu2024conformal,
  title={Robust and scalable uncertainty estimation with conformal prediction for machine-learned interatomic potentials},
  author={Hu, Yuge and Musielewicz, Joseph and Ulissi, Zachary W and Medford, Andrew J},
  journal={Machine Learning: Science and Technology},
  year={2024}
}

@article{tan2023single,
  title={Single-model uncertainty quantification in neural network potentials does not consistently outperform model ensembles},
  author={Tan, Aik Rui and Urata, Shingo and Goldman, Samuel and Dietschreit, Johannes CB and G{\'o}mez-Bombarelli, Rafael},
  journal={npj Computational Materials},
  volume={9},
  number={1},
  pages={225},
  year={2023},
  publisher={Nature Publishing Group UK London}
}

@article{wood2026family,
  title={UMA: A family of universal models for atoms},
  author={Wood, Brandon and Dzamba, Misko and Fu, Xiang and Gao, Meng and Shuaibi, Muhammed and Barroso-Luque, Luis and Abdelmaqsoud, Kareem and Gharakhanyan, Vahe and Kitchin, John and Levine, Daniel and others},
  journal={Advances in Neural Information Processing Systems},
  volume={38},
  pages={129391--129427},
  year={2026}
}

@article{neumann2024orb,
  title={Orb: A fast, scalable neural network potential},
  author={Neumann, Mark and Gin, James and Rhodes, Benjamin and Bennett, Steven and Li, Zhiyi and Choubisa, Hitarth and Hussey, Arthur and Godwin, Jonathan},
  journal={arXiv preprint arXiv:2410.22570},
  year={2024}
}

@inproceedings{graves2013speech,
  title={Speech recognition with deep recurrent neural networks},
  author={Graves, Alex and Mohamed, Abdel-rahman and Hinton, Geoffrey},
  booktitle={2013 IEEE international conference on acoustics, speech and signal processing},
  pages={6645--6649},
  year={2013},
  organization={Ieee}
}

@article{bishop1995training,
  title={Training with noise is equivalent to Tikhonov regularization},
  author={Bishop, Chris M},
  journal={Neural computation},
  volume={7},
  number={1},
  pages={108--116},
  year={1995},
  publisher={MIT Press}
}

@article{fuchs2020se,
  title={Se (3)-transformers: 3d roto-translation equivariant attention networks},
  author={Fuchs, Fabian and Worrall, Daniel and Fischer, Volker and Welling, Max},
  journal={Advances in neural information processing systems},
  volume={33},
  pages={1970--1981},
  year={2020}
}
\bibliographystyle{icml2026}

\newpage
\appendix
\onecolumn

\section{N-body Parameter Count: P-EGNN vs.\ Ensemble-EGNN}
\label{app:nbody_params}

We compare the parameter cost of a $K$-member EGNN deep ensemble against P-EGNN on the N-body benchmark. Both share the velocity-aware EGNN of \citet{satorras2021n} with $L\!=\!4$ layers and hidden width $d_h\!=\!64$; P-EGNN uses noise dimension $d_z\!=\!32$.

Each EGNN layer contains four two-linear-layer MLPs ($\phi_e,\phi_h,\phi_x,\phi_v$) with input widths $\{2d_h{+}3,\,2d_h,\,d_h,\,d_h\}$ and output widths $\{d_h,d_h,1,1\}$, where the $+3$ in $\phi_e$ accounts for the squared pairwise distance, the Euclidean pairwise distance, and the pairwise charge product. Including the charge-to-feature embedding, the full backbone has
\[
P \;=\; L\bigl(p_{\phi_e}+p_{\phi_h}+p_{\phi_x}+p_{\phi_v}\bigr) + p_{\mathrm{emb}} \;\approx\; 1.34\!\times\!10^5
\]
parameters, and a $K$-ensemble therefore costs $K\!\cdot\!P$.

P-EGNN augments a single backbone with the shared noise generator $W_z\!\in\!\mathbb{R}^{d_z\times d_z}$ (\cref{eq:noise_gen}) and a per-block projection $W^{\mathrm{noise}}_b\!\in\!\mathbb{R}^{d_h\times d_z}$ at the first linear layer of each perturbed block (\cref{eq:inject}). Per-sample E($n$)-equivariance restricts perturbations to the invariant scalar blocks $\phi_e$ and $\phi_h$, giving $B\!=\!2L\!=\!8$, so the overhead reduces to
\[
\Delta P \;=\; d_z^2 + B\,d_h\,d_z \;=\; 1{,}024 + 16{,}384 \;\approx\; 1.74\!\times\!10^4,
\]
i.e.\ ${\sim}13\%$ of $P$ and independent of the inference ensemble size $K$.

\begin{center}
\begin{tabular}{lrr}
\toprule
Model & Parameters & Ratio \\
\midrule
EGNN                      & $1.34\!\times\!10^5$ & $1.00\times$ \\
EGNN ensemble ($K\!=\!3$) & $4.02\!\times\!10^5$ & $3.00\times$ \\
P-EGNN (Ours)             & $1.51\!\times\!10^5$ & $1.13\times$ \\
\bottomrule
\end{tabular}
\end{center}

\end{document}